\documentclass{JHEP3}

\usepackage{amsmath,amssymb,amsfonts,mathrsfs}
\usepackage{url}
\usepackage{graphicx}
\usepackage[all]{xy}
\usepackage[vcentermath]{youngtab}
\usepackage{young}

\def\H{{\cal H}}

\title{Matrix Model and Refined Wall-Crossing Formula}

  \author{Haitao Liu\\
  Department of Mathematics and Statistics,\\
  University of New Brunswick, Fredericton, Canada, E3B 5A3\\
  and\\
  Theoretical Physics Group, The Blackett Laboratory,\\
   Imperial College London, London, UK\\
  and\\
  Department of Applied Mathematics,  Hebei University of Technology, Tianjin, China\\
  Email: \email{haitao.liu@unb.ca}}
  \author{Jie Yang\\
  School of Mathematical Sciences, Capital Normal University, Beijing, China\\
  and\\
  INFN, Sezione di Trieste, via Bonomea 265, Trieste, Italy\\
  Email: \email{yang9602@gmail.com} }
 \author{Jian Zhao\\
  SISSA and INFN, Sezione di Trieste, via Bonomea 265, Trieste, Italy\\
  Email: \email{zhaojian@sissa.it}}

\abstract{In this paper, we show how to get matrix models
corresponding to the refined BPS states partition functions of
$\mathbb{C}^3$, resolved conifold and $\mathbb{C}^3/\mathbb{Z}_2$ by
inserting the identity operator at a proper position in the
fermionic expression of the refined BPS states partition functions.
}


\preprint{}

\begin{document}
\section{Introduction}
Lately there has been progress in understanding the space of BPS
states, $\mathcal{H}_{BPS}$, in type IIA string compactifications on
Calabi-Yau threefolds. In general, such compactifications give rise
to the effective $\mathcal{N}=2$ theories in four dimensions.
$\mathcal{H}_{BPS}$ is a special subspace of the full Hilbert space
which is the one-particle representation of the $d=4, \mathcal{N}=2$
supersymmetry algebra. It contains lots of information about the
Calabi-Yau threefold $X$ and can be viewed as a bridge connecting
the black hole physics and topological strings \cite{Ooguri:2004zv}.

Due to the existence of the universal hypermultiplets,
$\mathcal{H}_{BPS}(\gamma)$ has the following decomposition
\begin{equation}
\mathcal{H}_{BPS}(\gamma)=(\mathbf{0},\mathbf{0};\mathbf{\frac12})\otimes
\mathcal{H}'_{BPS}(\gamma),
\end{equation}
where $\gamma$ is given by the generalized Mukai vector of the
stable coherent sheaves corresponding to the D6/D4/D2/D0 branes
\begin{equation}
\begin{array}{ccccccccc}
\gamma=ch(\mathcal{E})\sqrt{\hat{A}(X)}&=&p^0&+&P&+&Q&+&q_0\\
&\in& \text{H}^0&\oplus&\text{H}^2&\oplus&\text{H}^4&\oplus &\text{H}^6\\
& & D6 & &D4& &D2& &D0
\end{array}
\end{equation}

It is well known that the space $\mathcal{H}'_{BPS}$ depends on the
asymptotic boundary conditions in the four-dimensional spacetime,
where the boundary conditions in IIA compactification are the
complexified K\"{a}hler moduli $u=iJ+B$ of the Calabi-Yau threefold
$X$ \cite{Denef2007}. Roughly speaking, $\mathcal{H}'_{BPS}(\gamma,
u)\sim \text{H}^*(\mathcal{M}(\gamma,u))$, where
$\mathcal{M}(\gamma,u)$ is the moduli space of stable coherent
sheaves with the generalized Mukai vector $\gamma$ under certain
$u$-dependent stability condition \cite{Diaconescu2007}. The Spin(3)
action on $\mathcal{H}_{BPS}$ gives rise to the following refined
index of  $\mathcal{H}'_{BPS}$ \cite{Dimofte2010a}, after
factorizing the  contribution of the universal hypermultiplets,
\begin{equation}
\Omega^{ref}(\gamma, u, y):={\rm Tr}_{\H_{BPS}'(\gamma,u)}
(-y)^{2J_3'},
\end{equation}
where $J_3'$ is the reduced angular momentum \cite{Denef2007}.
$\Omega(\gamma, u, y)$ is conjectured to be related to the
Poincar\'{e} polynomial of the BPS states moduli space
\cite{Dimofte2010a}. Like the unrefined case we may define the
refined BPS states partition function \cite{Dimofte2010a} by
\begin{equation}
Z^{ref}_{BPS}(q,Q,y,u):=\sum_{\substack{\beta\in\text{H}_2(X;\mathbb{Z})\\
n\in\mathbb{Z}}}(-q)^nQ^\beta\Omega^{ref}(\gamma_{\beta,n},u,y).
\end{equation}
In \cite{Liu2010}, we have shown how to use the vertex operators in
2d free fermions and the crystal corresponding to the Calabi-Yau
threefold $X$ to reproduce the wall-crossing formula of the refined
BPS states partition function. In \cite{Liu2010}, we also conjecture
that for the toric CY without any compact four-cycles we have the
following formulas
\begin{equation}
Z_{BPS}^{ref}(q_1,q_2,Q)|_{chamber}=Z_{top}^{ref}(q_1,q_2,Q)Z_{top}^{ref}(q_1,q_2,Q^{-1})|_{chamber}.
\end{equation}

In this paper, we present a connection between the matrix model
with the $Z^{ref}_{BPS}$ by employing the method in
\cite{Ooguri2010a} to insert the identity operator at a proper
position to get a one-matrix model corresponding to the refined BPS
states partition function. In section \ref{sec:matrix} we review the work of
\cite{Ooguri2010a}. In section \ref{sec:refined_matrix} we show how to get the matrix
model corresponding to the refined BPS states partition function. In
section \ref{sec:conclusion} we give the summary and discussion on future research
directions.

\section{Matrix model and wall-crossing formula}
\label{sec:matrix}

\begin{figure}[ht]
\centering
\includegraphics[scale=0.3]{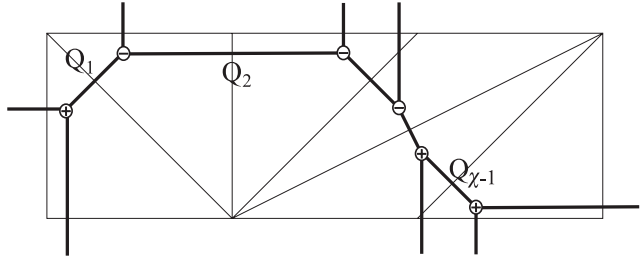}%
\caption{Toric diagram for Calabi-Yau threefold without compact four-cycles arises from a triangulation of a strip copied from \cite{Ooguri2010a}.}%
\label{toric}
\end{figure}

In this section we will review the matrix model for three
dimensional toric Calabi-Yau geometry without any compact four-cycles arising from a triangulation of a strip \cite{Sulkowski:2009rw, Ooguri2010a}. Let us denote the
Euler characteristic of the Calabi-Yau as $\chi$. Then
the number of base $\mathbb{P}^1$ of a toric CY 3-fold will be $\chi-1$ (see figure \ref{toric}).

We may define the following
creation and
annihilation operators by using the vertex operator in the 2d free fermions \cite{Sulkowski:2009rw, Young2008}:
\begin{equation}
  \label{eq:71}
A_-(x)\text{:=}\prod _{i=1}^{\chi } \Gamma _-^{s_i}\left(x \prod _{j=0}^{i-1} q_j\right),
\end{equation}
and
\begin{equation}
  \label{eq:3}
A_+(x)\text{:=}\prod _{i=1}^{\chi } \Gamma _+^{s_i}\left(x q \prod _{j=0}^{i-1} q_j^{-1}\right),
\end{equation}
where $s_i=1$ or $-1$,  and $q$ is defined in terms of the eigenvalues  $q_i$ of all color operators as
\begin{equation}
  \label{eq:73}
  q\text{:=}\prod _{i=0}^{\chi -1} q_i.
\end{equation}
The convention of $\Gamma$ matrices we will  use is
\begin{equation}
  \label{eq:72}
  \Gamma _{\pm }^{s_{i}=+1}(x)=\Gamma_\pm(x) ,\quad\quad \Gamma _{\pm }^{s_{i}=-1}(x)=\Gamma'_\pm(x),
\end{equation}
where the vertex operators $\Gamma$ are derived from two dimensional free fermion
theory \cite{Sulkowski:2009rw, Young2008} and they satisfy the following
commutation relation:
\begin{equation}
\label{Gammas} \Gamma _+^{s_1}(x)\Gamma _-^{s_2}(y)=\left(1-s_1s_2 x
y\right){}^{-s_1s_2}\Gamma _-^{s_2}(y)\Gamma _+^{s_1}(x).
\end{equation}

In terms of free fermions, the BPS partition functions can be
expressed as correlation functions of the vertex operators in 2d free fermions \cite{Sulkowski:2009rw}. The ket and
bra states of the NCDT chamber are generated by the creation and
annihilation operators as follows:
\begin{equation}
  \label{eq:1}
|\Omega _-\rangle := \prod _{r=0}^{\infty } A_-(q^r)|0\rangle, \quad\quad \langle \Omega _+| := \langle 0|\underset{l=0}{\overset{\infty }{\prod}}A_+(q^l).
\end{equation}
Therefore the partition function for the NCDT chamber is
\begin{equation}
  \label{eq:74}
  Z=\left\langle \Omega _+|\Omega _-\right\rangle.
\end{equation}
The corresponding matrix model partition function is obtained by inserting
the identity operator $\mathbb{I}$ of Hermitian matrix models, namely
\begin{equation}
  \label{eq:77}
  Z_{\rm matrix}=\langle\Omega_+|{\mathbb I}|\Omega_-\rangle,
\end{equation}
where
\begin{equation}
  \label{eq:identity}
\mathbb{I}= \int dU \left( \left.\prod _{i=1}^\infty \Gamma'_-(u_i) \right|0\right\rangle \left\langle 0\left|\prod _{j=1}^\infty \Gamma' _+(u_j^{-1})\right. \right).
\end{equation}
Here $dU$ is the unitary measure for $U(\infty)$ and $u_{i}=e^{i\phi_i}$ are the
eigenvalues of $U$:
\begin{equation}
dU=\prod_kd\phi_i\prod_{i<j}(e^{i\phi_i}-e^{i\phi_j})(e^{-i\phi_i}-e^{-i\phi_j}).
\end{equation}

\subsection{Matrix model for $\mathbb{C}^3$}
\begin{figure}[ht]
\centering
\includegraphics[scale=0.2]{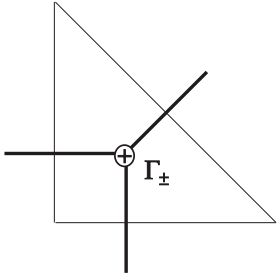}%
\caption{Toric diagram for $\mathbb{C}^3$.}%
\label{C3toric}
\end{figure}

The toric diagram of $\mathbb{C}^3$ is shown in figure \ref{C3toric}.

According to \cite{Sulkowski:2009rw}, we may define $A_-(x):=\Gamma _-^s(x)$ and
$A_+(x):=\Gamma _+^s(x q) $.\footnote{Here $s$ can be chosen to be
either $+1$ or $-1$ and the final two results may be connected by
an analytic continuation \cite{Ooguri2010a}.} We can split the integrand of the matrix
model partition function into
\begin{equation}
  \label{eq:5}
\left\langle 0\left|\underset{l=0}{\overset{\infty }{\prod
}}A_+\left(q^l\right)\prod _{i=1}^\infty \Gamma' _-(u_i)
\right|0\right\rangle=\prod _{i=1}^\infty \prod _{l=0}^{\infty }
\left(1+s u_i q^{l+1}\right)^s,
\end{equation}
and
\begin{equation}
  \label{eq:4}
\left\langle 0\left|\underset{j=1}{\overset{\infty }{\prod }} \Gamma'
_+(u_j^{-1})\prod _{r=0}^{\infty } A_-(q^r)\right|0\right\rangle
=\prod _{j=1}^\infty\prod _{r=0}^{\infty } \left(1+s
u_j^{-1}q^{r}\right)^s.
\end{equation}

Therefore the integrand of the matrix model is
\begin{equation}
  \label{eq:27}
\prod _{j=1}^\infty \prod _{r=0}^{\infty } \left(1+s u_j^{-1}q^{r}\right)^s\prod _{i=1}^\infty \prod _{l=0}^{\infty } \left(1+s u_i q^{l+1}\right)^s
=\left\{ \begin{array}{ll}
 {\rm Det}\displaystyle{\left(\prod _{r=0}^{\infty } \left(1+U^{-1} q^{r}\right)\left(1+U q^{r+1}\right)\right)}, & s=1 \\
\\
 {\rm Det}^{-1}\displaystyle{\left(\prod _{r=0}^{\infty } \left(1-U^{-1} q^{r}\right)\left(1-U q^{r+1}\right)\right)}, & s=-1
\end{array}
\right.  .
\end{equation}
where $U\in U(\infty)$ whose eigenvalues are $u_i$.

\subsection{Matrix model for the resolved conifold}
\begin{figure}[ht]
\centering
\includegraphics[scale=0.2]{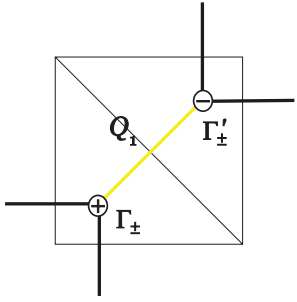}%
\caption{Toric diagram for the resolved conifold $\mathcal{O}(-1)\oplus\mathcal{O}(-1)\longrightarrow \mathbb{P}^1$.}%
\label{conifoldtoric}
\end{figure}

Figure  \ref{conifoldtoric} is the toric diagram for the resolved conifold $\mathcal{O}(-1)\oplus\mathcal{O}(-1)\longrightarrow \mathbb{P}^1$. According to the discussion of the vertex on a strip
\cite{Iqbal2006}, two $\mathbb{C}^3$ are connected by a $(-1,-1)$
curve, thus  $s_2=-s_1$.  Therefore we can choose $s_1=s$ and
$s_2=-s$. In the NCDT
chamber we denote $q=q_0q_1$, and $Q=-q_1$. Thus we can produce the BPS partition function by counting the pyramid model \cite{Szendroi2007, Sulkowski:2009rw, Yamazaki2010}.

According to \cite{Sulkowski:2009rw}, in the NCDT chamber, the creation operator and the annihilation operator are defined as follows:
\begin{equation}
  \label{eq:28}
  A_-(x)\text{:=}\prod _{i=1}^{\chi } \Gamma _-^{s_i}\left(x \prod _{j=0}^{i-1} q_j\right)=\Gamma _-^s\left(x q_0\right)\Gamma _-^{-s}(x q),
\end{equation}
and
\begin{equation}
  \label{eq:29}
  A_+(x)\text{:=}\prod _{i=1}^{\chi } \Gamma _+^{s_i}\left(x q \prod _{j=0}^{i-1} q_j^{-1}\right)=\Gamma _+^s\left(x q_1\right)\Gamma _+^{-s}(x).
\end{equation}
After inserting the matrix identity $\mathbb I$ we get two matrix
elements
\begin{equation}
  \label{eq:6}
\left\langle 0\left|\prod _{j=1}^\infty \Gamma' _+\left(u_j^{-1}\right)\prod _{r=0}^{\infty }
A_-\left(q^r\right)\right|0\right\rangle =\prod _{j=1}^\infty \prod _{r=0}^{\infty } \left(1+s u_j^{-1} q^rq_0\right)^s\left(1-s u_j^{-1}q^{r+1}\right)^{-s} ,
\end{equation}
and
\begin{equation}
  \label{eq:30}
  \left\langle 0\left|\underset{l=0}{\overset{\infty }{\prod }}A_+\left(q^l\right)\prod _{i=1}^\infty \Gamma' _-\left(u_i\right) \right|0\right\rangle
=\prod _{i=1}^\infty \prod _{l=0}^{\infty } \left(1-s u_iq^l\right){}^{-s}\left(1+s u_i q^l q_1\right){}^s.
\end{equation}
Then the partition function of the matrix model in the NCDT chamber is
\begin{equation}
  \label{eq:31}
\int dU  \text{Det}^s\left(\prod _{k=1}^{\infty }\frac{\left(1-s U^{-1} q^{k+1}Q^{-1}\right)\left(1-s U q^kQ\right)}
{\left(1-s U^{-1}q^{k+1}\right)\left(1-sU q^k\right)}\right).
\end{equation}
For chamber $R>0$, $0<n<B<n+1$ the wall crossing operator  is defined in \cite{Sulkowski:2009rw} as
\begin{equation}
    \overline{W}_{p=1}(x)=\Gamma _-^s(x)\hat{Q}_1\Gamma _+^{-s}(x)\hat{Q}_0.
\end{equation}
Therefore the partition function of the matrix model in the chamber $(R>0, 0<n<B<n+1)$ is
\begin{eqnarray}
  %\label{eq:53}
  Z_{BPS}|_{chamber\ n}&=&\left\langle\Omega _+\left|\mathbb{I}\left(\overline{W}_1(1)\right)^n\right|\Omega _-\right\rangle\nonumber\\
 &=&\int dU\left\langle \Omega _+\left|
\prod_{i=1}^\infty \Gamma'_-\left(u_i\right) \right|0\right\rangle \left\langle 0\left|\prod _{j=1}^\infty \Gamma'_+\left(u_j^{-1}\right)\left(\overline{W}_1(1)\right)^n\right|\Omega _-\right\rangle.
\end{eqnarray}

There are ambiguities of the position of the matrix identity
operator which will result in different partition functions of the matrix model. We will
discuss this problem in the final section.

We defined $\overline{W}'_{p=1}(x)$ by
\begin{equation}
  \label{eq:51}
  \overline{W}'_{p=1}(x):=\Gamma _+^s(x)\hat{Q}_1\Gamma _-^{-s}(x)\hat{Q}_0.
\end{equation}
In short we list the partition functions $Z_{n|p}$ for all chambers in the resolved conifold  ($n\geq0$)

\begin{table}[h]
  \centering
   \begin{tabular}{lll}
    $R>0, B\in [n, n+1]$ & \quad\quad $Z_{n|1}=\langle \Omega_+|\overline W^n_1|\Omega_-\rangle$ ,
\\
    $R>0, B\in [-n-1, -n]$ & \quad\quad $Z'_{n+1|1}=\langle\Omega_+|(\overline W'_1)^{n+1}|\Omega_-\rangle$ ,
\\
   $R<0, B\in [n, n+1]$ & \quad\quad $\widetilde Z_{n+1|1}=\langle 0|\overline W^{n+1}_1|0\rangle$ ,
\\
    $R<0, B\in [-n-1,-n]$ & \quad\quad $\widetilde Z'_{n|1}=\langle0|(\overline W'_1)^n|0\rangle$ .
\end{tabular}
\end{table}

The corresponding matrix models are the results of the insertion of the identity operator in the partition functions respectively.

\subsection{Matrix model for $\mathbb{C}^3/\mathbb{Z}_2$}

\begin{figure}[ht]
\centering
\includegraphics[scale=0.2]{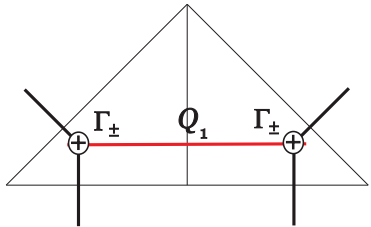}%
\caption{Toric diagram for the resolved $\mathbb{C}^3/\mathbb{Z}_2$.}%
\label{C3Z2}
\end{figure}
Figure \ref{C3Z2} is the toric diagram of $\mathcal{O}_{\mathbb{P}^1}(-2, 0)$ which is the resolved $\mathbb{C}^3/\mathbb{Z}_2$. The vertex strip in this case is different from the resolved conifold by $s_1=s_2$ rather than $s_1=-s_2$.  In the NCDT chamber, we define the creation operator
\begin{equation}
  A_-(x)\text{:=}\prod _{i=1}^{\chi } \Gamma _-^{s_i}\left(x \prod _{j=0}^{i-1} q_j\right)=\Gamma _-^s\left(x q_0\right)\Gamma _-^{s}(x q),
\end{equation}
and the annihilation operator
\begin{equation}
  A_+(x)\text{:=}\prod _{i=1}^{\chi } \Gamma _+^{s_i}\left(x q \prod _{j=0}^{i-1} q_j^{-1}\right)=\Gamma _+^s\left(x q_1\right)\Gamma _+^{s}(x).
\end{equation}
The insertion of the matrix identity $\mathbb I$ results in two matrix elements
\begin{equation}
\left\langle 0\left|\prod _{j=1}^\infty \Gamma' _+\left(u_j^{-1}\right)\prod _{r=0}^{\infty }
A_-\left(q^r\right)\right|0\right\rangle =\prod _{j=1}^\infty \prod _{r=0}^{\infty } \left(1+s u_j^{-1} q^rq_0\right)^s\left(1+s u_j^{-1}q^{r+1}\right)^{s} ,
\end{equation}
and
\begin{equation}
  \left\langle 0\left|\underset{l=0}{\overset{\infty }{\prod }}A_+\left(q^l\right)\prod _{i=1}^\infty \Gamma' _-\left(u_i\right) \right|0\right\rangle
=\prod _{i=1}^\infty \prod _{l=0}^{\infty } \left(1+s u_iq^l\right){}^{s}\left(1+s u_i q^l q_1\right){}^s.
\end{equation}
Then the partition function of the matrix model is
\begin{equation}
\int dU  \text{Det}^s\left(\prod _{k=1}^{\infty }\left(1-s U^{-1} q^{k+1}Q^{-1}\right)\left(1-s U q^kQ\right)
\left(1+s U^{-1}q^{k+1}\right)\left(1+sU q^k\right)\right).
\end{equation}
The wall crossing operator is defined in \cite{Sulkowski:2009rw} as
\begin{equation}
  \overline{W}_1(x)=\Gamma _+^s(x)\hat{Q}_1\Gamma _-^{s}(x)\hat{Q}_0.
\end{equation}
Therefore the partition function of the matrix model for chamber  $R>0$, $0<n<B<n+1$  is
\begin{equation}
  %\label{eq:53}
  Z_{BPS}|_{chamber\ n}=\int dU\left\langle \Omega _+\left|
\prod_{i=1}^N \Gamma'_-\left(u_i\right) \right|0\right\rangle \left\langle 0\left|\prod _{j=1}^N \Gamma'_+\left(u_j^{-1}\right)\left(\overline{W}_1(1)\right)^n\right|\Omega _-\right\rangle.
\end{equation}
Similarly we can get the matrix models in all the chambers as previous section.

\section{refined Matrix model and refined wall-crossing formula}
\label{sec:refined_matrix}

In this section, we will use techniques introduced in previous section
and the refined BPS states partition functions proposed in
\cite{Liu2010} to obtain refined matrix model for several typical
toric Calabi-Yau 3-folds.

\subsection{Refined matrix model for $\mathbb{C}^3$}

In \cite{Liu2010}, we define the creation and annihilation operators as
\begin{equation}
  \label{eq:26}
  \overline{A}_-(x)\text{:=}\hat{Q}_{0,-}^{ \frac12 -\frac{\delta}{2}}\Gamma _-(x)\hat{Q}_{0,-}^{ \frac12 +\frac{\delta}{2}}=\Gamma_-\left(x q_2^{ \frac12 -\frac{\delta}{2}}\right)\hat{Q}_{0,-},
\end{equation}
\begin{equation}
  \label{eq:79}
  \overline{A}_+(x)\text{:=}\hat{Q}_{0,+}^{ \frac12 -\frac{\delta}{2}}\Gamma _+(x)\hat{Q}_{0,+}^{ \frac12 +\frac{\delta}{2}}
=\hat{Q}_{0,+}\Gamma_+\left(x q_1^{ \frac12 +\frac{\delta}{2}}\right),
\end{equation}
and states
\begin{equation}
\langle\Omega_+|:=\langle0| \overline A_+(1) \cdots \overline A_+(1)=\langle0|\prod_{i=1}^\infty\Gamma_+(q_1^{i-\frac{1}{2}+\frac{\delta}{2}}),
\end{equation}
\begin{equation}
|\Omega_-\rangle:= \overline A_-(1) \cdots \overline A_-(1)|0\rangle=\prod_{j=1}^\infty\Gamma_-(q_2^{j-\frac{1}{2}-\frac\delta2})|0\rangle.
\end{equation}
In the general convention, we can rewrite  $\langle\Omega^s_+|$ and $|\Omega^s_-\rangle$  as follows
\begin{eqnarray}
\langle\Omega^s_+|&:=&\langle0|\prod_{i=1}^\infty\Gamma^s_+(q_1^{i-\frac{1}{2}+\frac{\delta}{2}}) =\left\{\begin{array}{lll}
\langle0|\prod_{i=1}^\infty\Gamma_+(q_1^{i-\frac{1}{2}+\frac{\delta}{2}}) \ \ \ \text{ if\ } s=1,\\
\langle0|\prod_{i=1}^\infty\Gamma'_+(q_1^{i-\frac{1}{2}+\frac{\delta}{2}}) \ \ \ \text{ if\ } s=-1,
\end{array}\right.\\
|\Omega^s_-\rangle&:=&\prod_{j=1}^\infty\Gamma^s_-(q_2^{j-\frac{1}{2}-\frac\delta2})|0\rangle=\left\{\begin{array}{lll}
\prod_{j=1}^\infty\Gamma_-(q_2^{j-\frac{1}{2}-\frac\delta2})|0\rangle \quad  \ \text{ if\ } s=1,\\
\prod_{j=1}^\infty\Gamma'_-(q_2^{j-\frac{1}{2}-\frac\delta2})|0\rangle \quad \ \text{ if\ } s=-1.
\end{array}\right.
\end{eqnarray}
Then the refined BPS states partition function is
\begin{equation}
\mathcal{Z}_{BPS}^{ref}=\langle\Omega^s_+|\Omega^s_-\rangle=M_\delta(q_1, q_2)
\label{eq:ref_C3}
\end{equation}
where the refined MacMahon function $M_\delta(q_1, q_2)$ is defined by
\begin{equation}
M_\delta(q_1, q_2)=\prod_{i,j=1}^\infty(1-q_1^{i-\frac{1}{2}+\frac{\delta}{2}}q_2^{j-\frac{1}{2}-\frac{\delta}{2}})^{-1}.
\end{equation}
In order to get a matrix model we insert the identity operator $\mathbb{I}$ into the formula (\ref{eq:ref_C3})
\begin{equation}
\mathcal{Z}_{BPS}^{ref}=\left\langle0\left|\prod_{i=1}^\infty\Gamma^s_+(q_1^{i-\frac{i}{2}+\frac{\delta}{2}}) \ \mathbb{I}\
\prod_{j=1}^\infty\Gamma^s_-(q_2^{j-\frac{1}{2}-\frac\delta2})\right|0\right\rangle=M_\delta(q_1, q_2).
\end{equation}
Due to the formula (\ref{eq:identity}) of $\mathbb{I}$,
the matrix elements are
\begin{eqnarray}
  \label{eq:7}
\left\langle 0\left|\prod _{j=1}^\infty \Gamma'_+\left(u_j^{-1}\right)\right|\Omega^s _-\right\rangle
&=&\left\langle 0\left|\prod _{j=1}^\infty \Gamma'_+\left(u_j^{-1}\right)\prod_{k=1}^\infty\Gamma^s_-\left(q_2^{k- \frac12 -\frac{\delta}{2}}\right)\right|0\right\rangle \nonumber\\
 &=&\prod _{k=1}^{\infty} \prod _{j=1}^\infty \left(1+ su_j^{-1}q_2^{k- \frac12 -\frac{\delta}{2}}\right)^s ,
\end{eqnarray}
and
\begin{eqnarray}
  \label{eq:8}
\left\langle \Omega^s_+\left|\prod _{i=1}^\infty \Gamma'_-(u_i) \right|0\right\rangle
&=&\left\langle 0\left|\prod_{l=1}^\infty\Gamma^s_+\left(q_1^{l-\frac12+\frac\delta2}\right)\prod_{i=1}^\infty \Gamma'_-(u_i) \right|0\right\rangle \nonumber\\
&=&\prod _{l=1}^{\infty } \prod_{i=1}^\infty \left(1+ s u_i q_1^{l-\frac12 +\frac{\delta}{2}}\right)^s  .
\end{eqnarray}
Therefore the matrix model integrand is
\begin{equation}
  \label{eq:39}
  \text{Det}^{s}\left[\prod _{k=1}^{\infty } \left(1+ sU^{-1}q_2^{k- \frac12 -\frac{\delta}{2}}\right)\left(1+ sU q_1^{k- \frac12 +\frac{\delta}{2}}\right)\right].
\end{equation}
Finally we have
\begin{equation}
\mathcal{Z}_{BPS}^{ref}=\int dU \text{Det}^{s}\left[\prod _{k=1}^{\infty } \left(1+ sU^{-1}q_2^{k- \frac12 -\frac{\delta}{2}}\right)\left(1+ sU q_1^{k- \frac12 +\frac{\delta}{2}}\right)\right],
\end{equation}
where $dU$ denotes the unitary measure for $U(\infty)$. It is given by
\begin{equation}
dU=\prod_kd\phi_k\prod_{i<j}(e^{i\phi_i}-e^{i\phi_j})(e^{-i\phi_i}-e^{-i\phi_j}),
\end{equation}
where $u_i=e^{i\phi_i}$ are the eigenvalues of $U$.

\subsection{Refined matrix model for the resolved conifold}

We follow the same logic as the $\mathbb{C}^3$ case. First we define $\langle\Omega^s_+|$ and $|\Omega^s_ -\rangle$ by
\begin{eqnarray}
  |\Omega^s_-\rangle&:=&\prod_{j=1}^\infty\Gamma^s_-\left(q_{2}^{j- \frac12 }(-Q)^{- \frac12 }\right)\Gamma^{-s}_-\left(q_{2}^{j-\frac12 }(-Q)^{ \frac12 }\right)|0\rangle,   \label{eq:40}\\
  \langle\Omega^s_+|&:=&\langle0|\prod_{i=1}^\infty\Gamma^s_+\left(q_{1}^{i-\frac12 }(-Q)^{ \frac12 }\right)\Gamma^{-s}_+\left(q_{1}^{i-\frac12 }(-Q)^{- \frac12 }\right).  \label{eq:41}
\end{eqnarray}
Then the refined BPS states partition function in the NCDT chamber can be written as
\begin{eqnarray}
\left.\mathcal{Z}_{BPS}^{ref}\right|_{NCDT}&=&\langle\Omega^s_+|\Omega^s_-\rangle\nonumber\\
&=&\left(M_{\delta=0}(q_1, q_2)\right)^2\prod_{i,j=1}^\infty(1-q_1^{i-\frac12}q_2^{j-\frac12}Q)(1-q_1^{i-\frac12}q_2^{j-\frac12}Q^{-1}).\label{eq42}
\end{eqnarray}
Actually there are some degrees of freedom of the variables of $\Gamma$ operators in the formulas (\ref{eq:40}, \ref{eq:41}).
We find that in order to preserve the equation (\ref{eq42}), the general definition of $\langle\Omega^s_+|$ and $|\Omega^s_-\rangle$ will be as follows
\begin{eqnarray}
  |\Omega^{(s,\delta_1,\delta_2)}_-\rangle&:=&\prod_{j=1}^\infty\Gamma^s_-\left(q_{2}^{j- \frac12+\delta_2 }q_1^{\delta_1}(-Q)^{- \frac12 }\right)\Gamma^{-s}_-\left(q_{2}^{j-\frac12+\delta_2 }q_1^{\delta_1}(-Q)^{ \frac12 }\right)|0\rangle,   \label{eq40}\\
  \langle\Omega^{(s,\delta_1,\delta_2)}_+|&:=&\langle0|\prod_{i=1}^\infty\Gamma^s_+\left(q_{1}^{i-\frac12-\delta_1 }q_2^{-\delta_2}(-Q)^{ \frac12 }\right)\Gamma^{-s}_+\left(q_{1}^{i-\frac12-\delta_1 }q_2^{-\delta_2}(-Q)^{- \frac12 }\right),  \label{eq41}
\end{eqnarray}
where $\delta_1, \delta_2$ are two arbitrary integers.
Now we insert the identity operator $\mathbb{I}$ as follows:
\begin{eqnarray}
\left.\mathcal{Z}_{BPS}^{ref}\right|_{NCDT}&=&\langle\Omega^{(s,\delta_1,\delta_2)}_+|\mathbb{I}|\Omega^{(s,\delta_1,\delta_2)}_-\rangle=\langle0|\prod_{i=1}^\infty\Gamma^s_+\left(q_{1}^{i-\frac12-\delta_1 }q_2^{-\delta_2}(-Q)^{ \frac12 }\right)\Gamma^{-s}_+\left(q_{1}^{i-\frac12-\delta_1 }q_2^{-\delta_2}(-Q)^{- \frac12 }\right)\cdot\nonumber\\
&&\mathbb{I}\prod_{j=1}^\infty\Gamma^s_-\left(q_{2}^{j- \frac12+\delta_2 }q_1^{\delta_1}(-Q)^{- \frac12 }\right)\Gamma^{-s}_-\left(q_{2}^{j-\frac12+\delta_2 }q_1^{\delta_1}(-Q)^{ \frac12 }\right)|0\rangle.
\end{eqnarray}
Thus we may obtain the following matrix elements
\begin{eqnarray}
  \label{eq:9}
&&\left\langle 0\left|\prod _{j=1}^\infty \Gamma'_+\left(u_j^{-1}\right)\prod_{r=1}^\infty\Gamma^s_-\left(q_{2}^{r- \frac12+\delta_2 }q_1^{\delta_1}(-Q)^{- \frac12 }\right)\Gamma^{-s}_-\left(q_{2}^{r-\frac12+\delta_2 }q_1^{\delta_1}(-Q)^{ \frac12 }\right)\right|0\right\rangle \nonumber\\
&=&\prod _{j=1}^\infty \prod _{r=1}^{\infty} \left(1+s u_j^{-1}(-Q)^{- \frac12} q_2^{r-\frac12+\delta_2}q_1^{\delta_1}\right)^s\left(1-su_j^{-1}(-Q)^{ \frac12}q_2^{r-\frac12+\delta_2}q_1^{\delta_1}\right)^{-s},
\end{eqnarray}
and
\begin{eqnarray}
  \label{eq:20}
&&\left\langle0\left|\prod_{l=1}^\infty\Gamma^s_+\left(q_{1}^{l-\frac12-\delta_1 }q_2^{-\delta_2}(-Q)^{ \frac12 }\right)\Gamma^{-s}_+\left(q_{1}^{l-\frac12-\delta_1 }q_2^{-\delta_2}(-Q)^{- \frac12 }\right)\prod _{i=1}^\infty \Gamma'_-(u_i) \right|0\right\rangle \nonumber\\
&=&\prod _{i=1}^\infty \prod _{l=1}^{\infty } \left(1+s u_i (-Q)^{\frac12 }q_1^{l-\frac12-\delta_1}q_2^{-\delta_2}\right)^s\left(1-su_i (-Q)^{-\frac{1}{2}} q_1^{l-\frac12-\delta_1}q_2^{-\delta_2}\right)^{-s}.
\end{eqnarray}

Then the partition of the matrix model is
\begin{equation}
  \label{eq:conifold}
\left.Z_{BPS}^{ref}\right|_{NCDT}=\int dU  \text{Det}^s\left[\prod _{k=0}^{\infty } \frac{\left(1+s U^{-1}(-Q)^{- \frac12} q_2^{k-\frac12+\delta_2}q_1^{\delta_1}\right)
 \left(1+s U (-Q)^{\frac12 }q_1^{k-\frac12-\delta_1}q_2^{-\delta_2}\right)}{\left(1-sU^{-1}(-Q)^{ \frac12}q_2^{k-\frac12+\delta_2}q_1^{\delta_1}\right)
\left(1-sU (-Q)^{ -\frac12} q_1^{k-\frac12-\delta_1}q_2^{-\delta_2}\right)}\right].
\end{equation}

For chamber $(R>0, 0<n<B<n+1)$, the partition function is
\begin{eqnarray}
  %\label{eq:59}
\left.Z_{BPS}^{ref}\right|_{(R>0, 0<n<B<n+1)}&=&  \left\langle 0\left|\prod _{k=1}^{\infty } \Gamma _+^s\left(q_1^{k- \frac12-\delta_1 }q_2^{-\delta_2}(-Q)^{ \frac12 }\right) \Gamma _+^{-s}
\left(q_1^{k+n- \frac12-\delta_1 }q_2^{-\delta_2}(-Q)^{- \frac12 }\right) \right.\right.\nonumber\\
&&\bullet\prod _{l=1}^n \Gamma _-^s\left(q_2^{l- \frac12+\delta_2 }q_1^{\delta_1}(-Q)^{- \frac12 }\right) \Gamma _+^{-s}\left(q_1^{n-l+ \frac12-\delta_1 }q_2^{-\delta_2}(-Q)^{- \frac12 }\right)   \nonumber\\
&&\left.\left.\bullet \prod _{k=1}^{\infty}  \Gamma _-^s\left(q_2^{k+n- \frac12+\delta_2 }q_1^{\delta_1}(-Q)^{- \frac12 }\right)
\Gamma_-^{-s}\left(q_2^{k- \frac12+\delta_2 }q_1^{\delta_1}(-Q)^{ \frac12 }\right) \right|0\right\rangle.\nonumber\\
&&
\end{eqnarray}
Now we insert the matrix identity operator $\mathbb I$ in the partition function and we show the details in appendix (\ref{app:coni}), then it gives rise to
\begin{eqnarray}
  %\label{eq:63}
  &&\left. Z^{ref}_{BPS}\right|_{(R>0, 0<n<B<n+1)}=\int dU\prod _{l=1}^n \prod _{p=1}^l \left(1-Q^{-1} q_1^{n-l+ \frac12 }q_2^{p- \frac12 }\right)^{-1}\nonumber\\
  &&\bullet\text{Det}^s\left[\prod _{k=0}^{\infty } \frac{\left(1+s U^{-1}(-Q)^{- \frac12} q_2^{r-\frac12+\delta_2}q_1^{\delta_1}\right)
 \left(1+s U (-Q)^{\frac12 }q_1^{l-\frac12-\delta_1}q_2^{-\delta_2}\right)}{\left(1-sU^{-1}(-Q)^{ \frac12}q_2^{r-\frac12+\delta_2}q_1^{\delta_1}\right)
\left(1-sU (-Q)^{ -\frac12} q_1^{l-\frac12-\delta_1}q_2^{-\delta_2}\right)}\right].
\end{eqnarray}
Similarly, the partition function of the matrix model for the chamber $(R>0, n-1<0<n\leq0)$ is
\begin{eqnarray}
  %\label{eq:63}
  &&\left. Z^{ref}_{BPS}\right|_{(R>0, n-1<B<n\leq0)}=\int dU\prod _{l=1}^n \prod _{p=1}^l \left(1-Q q_1^{n-l+ \frac12 }q_2^{p- \frac12 }\right)^{-1}\nonumber\\
  &&\bullet\text{Det}^s\left[\prod _{k=0}^{\infty } \frac{\left(1+s U^{-1}(-Q)^{\frac12} q_2^{r-\frac12+\delta_2}q_1^{\delta_1}\right)
 \left(1+s U (-Q)^{-\frac12 }q_1^{l-\frac12-\delta_1}q_2^{-\delta_2}\right)}{\left(1-sU^{-1}(-Q)^{ -\frac12}q_2^{r-\frac12+\delta_2}q_1^{\delta_1}\right)\left(1-sU (-Q)^{ \frac12} q_1^{l-\frac12-\delta_1}q_2^{-\delta_2}\right)}\right].
\end{eqnarray}

\subsection{Refined matrix model for $\mathbb{C}^3/\mathbb{Z}_2$}

According to \cite{Liu2010}, the refined BPS states partition function of $\mathbb{C}^3/\mathbb{Z}_2$ in the chamber $(R>0, 0\leq n<B<n+1)$ is
\begin{eqnarray}
&&\left.Z_{BPS}^{ref}\right|_{(R>0, 0\leq n<B<n+1)}\nonumber\\
&=&\langle0|\prod_{k=1}^\infty \Gamma_+^s\left[q_1^{k+\delta_1}q_2^{\delta_2}(-Q)^{\frac{1}{2}}\right]
\Gamma_+^s\left[q_1^{k+n+\delta_1}q_2^{\delta_2}(-Q)^{-\frac{1}{2}}\right] \Gamma_-^s\left[q_2^{-\delta_2}q_1^{-\delta_1}(-Q)^{-\frac{1}{2}}\right]\cdot \nonumber\\
& &\quad\times\Gamma_+^s\left[ q_1^{n+\delta_1}q_2^{\delta_2}(-Q)^{-\frac{1}{2}}\right] \Gamma_-^s\left[q_2^{1-\delta_2}q_1^{-\delta_1}(-Q)^{-\frac{1}{2}}\right]\Gamma_+^s\left[ q_1^{n-1+\delta_1}q_2^{\delta_2}(-Q)^{-\frac{1}{2}}\right]\cdots\nonumber\\
& &\quad\times\Gamma_-^s\left[q_2^{n-1-\delta_2}q_1^{-\delta_1}(-Q)^{-\frac{1}{2}}\right]\Gamma_+^s\left[ q_1^{1+\delta_1}q_2^{\delta_2}(-Q)^{-\frac{1}{2}}\right]\cdot \nonumber\\
&&\quad\times\prod_{k=1}^\infty\Gamma_-^s\left[q_2^{k+n-1-\delta_2}q_1^{-\delta_1}(-Q)^{-\frac{1}{2}}\right]\Gamma_-^s\left[q_2^{k-1-\delta_2}q_1^{-\delta_1}(-Q)^{\frac{1}{2}}\right]|0\rangle\nonumber\\
&=&\prod_{l+r\leq n+1}(1-q_1^{l}q_2^{r-1}Q^{-1})\cdot\langle0|\prod_{k=1}^\infty \Gamma_+^s\left[q_1^{k+\delta_1}q_2^{\delta_2}(-Q)^{\frac{1}{2}}\right]\Gamma_+^s\left[q_1^{k+\delta_1}q_2^{\delta_2}(-Q)^{-\frac{1}{2}}\right] \cdot \nonumber\\
&&\quad\times\prod_{k=1}^\infty\Gamma_-^s\left[q_2^{k-1-\delta_2}q_1^{-\delta_1}(-Q)^{-\frac{1}{2}}\right]\Gamma_-^s
\left[q_2^{k-1-\delta_2}q_1^{-\delta_1}(-Q)^{\frac{1}{2}}\right]|0\rangle\nonumber\\
&=& M_{\delta=\frac12}^2(q_1,q_2)\prod_{i,j=1}^\infty(1-q_1^{i}q_2^{j-1}Q)^{-1}\prod_{i+j>n+1}(1-q_1^{i}q_2^{j-1}Q^{-1})^{-1}
\end{eqnarray}

Now we insert the identity operator $\mathbb{I}$ as follows:
\begin{eqnarray}
&&\langle0|\prod_{k=1}^\infty \Gamma_+^s\left[q_1^{k+\delta_1}q_2^{\delta_2}(-Q)^{\frac{1}{2}}\right]\Gamma_+^s\left[q_1^{k+\delta_1}q_2^{\delta_2}(-Q)^{-\frac{1}{2}}\right] \cdot \mathbb{I}\nonumber\\
&&\prod_{k=1}^\infty\Gamma_-^s\left[q_2^{k-1-\delta_2}q_1^{-\delta_1}(-Q)^{-\frac{1}{2}}\right]\Gamma_-^s\left[q_2^{k-1-\delta_2}q_1^{-\delta_1}(-Q)^{\frac{1}{2}}\right]|0\rangle.
\end{eqnarray}
Then the matrix elements are
\begin{eqnarray}
&&\left\langle0\left|\prod_{k=1}^\infty \Gamma_+^s\left[q_1^{k+\delta_1}q_2^{\delta_2}(-Q)^{\frac{1}{2}}\right]\Gamma_+^s\left[q_1^{k+\delta_1}q_2^{\delta_2}(-Q)^{-\frac{1}{2}}\right]\prod_{i=1}^\infty\Gamma_-'(u_i) \right|0\right\rangle\nonumber\\
&=&\prod _{i=1}^\infty \prod _{k=1}^{\infty } \left(1+s u_i(-Q)^{-\frac12}q_1^{k+\delta_1}q_2^{\delta_2}\right)^s \left(1+s u_i (-Q)^{\frac12}q_1^{k+\delta_1}q_2^{\delta_2}\right)^s,
\end{eqnarray}
and
\begin{eqnarray}
&&\left\langle0\left|\prod_{j=1}^\infty\Gamma_+'\left(u_j^{-1}\right)\prod_{k=1}^\infty\Gamma_-^s\left[q_2^{k-1-\delta_2}q_1^{-\delta_1}(-Q)^{-\frac{1}{2}}\right]\Gamma_-^s\left[q_2^{k-1-\delta_2}q_1^{-\delta_1}(-Q)^{\frac{1}{2}}\right]\right|0\right\rangle\nonumber\\
&=&\prod _{j=1}^\infty \prod _{k=1}^{\infty } \left(1+s u_j^{-1} (-Q)^{-\frac12}q_2^{k-1-\delta_2}q_1^{-\delta_1}\right)^s \left(1+s u_j^{-1} (-Q)^{\frac12}q_2^{k-1-\delta_2}q_1^{-\delta_1}\right)^s.
\end{eqnarray}
Therefore the partition function of the matrix model for the $(R>0, 0\leq n<B<n+1)$ chamber is
\begin{eqnarray}
  \label{eq:47}
&&\left.Z_{BPS}^{ref}\right|_{(R>0, 0\leq n<B<n+1)}\nonumber\\
&=&\prod_{l+r\leq n+1}(1-q_1^{l}q_2^{r-1}Q^{-1})^{-1}\int dU \text{Det}^s\left[\prod _{k=1}^{\infty } \left(1+s U(-Q)^{-\frac12}q_1^{k+\delta_1}q_2^{\delta_2}\right)\left(1+s U (-Q)^{\frac12}q_1^{k+\delta_1}q_2^{\delta_2}\right) \right.\nonumber\\
&&\left.\times\left(1+s U^{-1} (-Q)^{-\frac12}q_2^{k-1-\delta_2}q_1^{-\delta_1}\right) \left(1+s U^{-1} (-Q)^{\frac12}q_2^{k-1-\delta_2}q_1^{-\delta_1}\right)\right].\nonumber\\
&&
\end{eqnarray}

\section{Conclusion and discussion}
\label{sec:conclusion}

In this paper, we use the free fermion version of refined BPS states partition
functions to obtain their corresponding matrix models. But there still are
some subtle problems hidden in calculations.

The first one is the
choice of $s$, the ``type'' of the first vertex in a strip. As argued
in \cite{Ooguri2010a}, the final results should have an analytic
continuation. Here we want to give some simple arguments on this analytic
continuation. The key formula in this paper is the equation (\ref{Gammas}), from
which we may see if we change one of the $s$'s into its opposite one,
then variables will go from numerator to denominator or from
denominator to numerator. This is similar as the analytic continuation
on $\mathbb{P}^1$, which has two patches and we can construct the analytic
continuation from one patch to the other.

Another subtle problem is
the choice of $\delta$ in the refined MacMahon function $M_{\delta}(q_1, q_2)$.
In fact $\delta$ is an arbitrary constant set up by hand if we just want to get the generating function
of 3d partition function. While it is not clear to us whether the
choice of $\delta$ in the paper \cite{RTV} is unique or not, how to get the refined MacMahon function appearing in \cite{Behrend2009} which gives
the mathematical rigid refined BPS partition
functions for the D0 branes, and whether those different refined MacMahon functions in \cite{Behrend2009} and \cite{Liu2010} are
physically identical.

The third subtle problem comes from the
position of insertion of the identity operator. Apparently,
a different inserting position will give rise to a different action of matrix
model. But just as in QFT, an identity operator means summation over
complete set of intermediate states, and inserting an identity operator at different positions just means we observe different
stages of interactions. Actually, if we want to get a multi-matrix
model rather than a one-matrix model we may insert more identity
operators in the corresponding correlation function of refined BPS
partition function.

In \cite{Ooguri2010a} besides getting the matrix models corresponding to the BPS partition function, the authors also find the following interesting property of the BPS partition function: the matrix model for the BPS counting on the CY $X$ is related to the topological string partition function for another CY $Y$, whose K\"{a}hler moduli space $\mathcal{M}(Y)$ contains two copies of $\mathcal{M}(X)$, e.g. the partition function of matrix model corresponding to the BPS partition function on the conifold will be related to the topological string partition function on the SPP geometry. It would be interesting to see if the matrix model proposed in this paper is related to the refined topological string partition function on another CY. This work is under consideration.

Along the line of techniques discussed in the paper, we can also
obtain refined matrix models for any strip like toric CY quickly,
and what's more, if we insert an identity in the equation (150) of \cite{RTV} we can
get a matrix model for the refined topological vertex. Since the refined
topological vertex is the element to generate 5d instanton partition
function, we can obtain a matrix model for $U(N)$ ${\cal N}=2$ instanton
partition function. But the matrix models obtained by using this method have too many
matrices and are very difficult to deal with.

As in \cite{Ooguri2010a}, in addition to inserting the identity operator, there is also another way to obtain matrix model from BPS partition
functions, namely the non-intersecting path method introduced in \cite{Eynard2009, Eynard2010a, Eynard2010, Ooguri2010a}. It would be interesting to see how to reproduce the matrix models presented in this paper by using non-intersecting path method and how to use this method to get matrix models of refined topological vertex. We hope that the investigation on the relationship between the non-intersecting paths and the refined topological vertex will deepen our understanding on the refined topological vertex and refined BPS partition function. This work is in progress.

\acknowledgments { H. Liu thanks the hospitality of the theoretical
physics group of Imperial College while preparing this paper and
also appreciates Professor Jack Gegenberg for his support. The
research of H. Liu is partially supported by the ``Pam and John
Little Overseas Scholarship'' in the University of New Brunswick,
Canada. J. Yang is supported by the Scientific Research Foundation
for The Excellent Youth Scholars of Capital Normal University,
Beijing. J. Zhao thanks G. Bonelli and  A. Tanzini for many useful
discussions in related topics and giving him much support in
research. }

\appendix
\section{Refined matrix model for different chambers of the resolved conifold }
\label{app:coni}
\subsection{The chamber $(R>0, 0<n<B<n+1)$}
The refined partition function for chamber $(R>0, 0<n<B<n+1)$  of the resolved conifold is
\begin{eqnarray}
  \label{eq:59}
\left.Z_{BPS}^{ref}\right|_{(R>0, 0<n<B<n+1)}&=&  \left\langle 0\left|\prod _{k=1}^{\infty } \Gamma _+^s\left(q_1^{k- \frac12-\delta_1 }q_2^{-\delta_2}(-Q)^{ \frac12 }\right) \Gamma _+^{-s}
\left(q_1^{k+n- \frac12-\delta_1 }q_2^{-\delta_2}(-Q)^{- \frac12 }\right) \right.\right.\nonumber\\
&&\bullet\prod _{l=1}^n \Gamma _-^s\left(q_2^{l- \frac12+\delta_2 }q_1^{\delta_1}(-Q)^{- \frac12 }\right) \Gamma _+^{-s}\left(q_1^{n-l+ \frac12-\delta_1 }q_2^{-\delta_2}(-Q)^{- \frac12 }\right)   \nonumber\\
&&\left.\left.\bullet \prod _{k=1}^{\infty}  \Gamma _-^s\left(q_2^{k+n- \frac12+\delta_2 }q_1^{\delta_1}(-Q)^{- \frac12 }\right)
\Gamma_-^{-s}\left(q_2^{k- \frac12+\delta_2 }q_1^{\delta_1}(-Q)^{ \frac12 }\right) \right|0\right\rangle.\nonumber\\
&&
\end{eqnarray}

We use the commutation relation of $\Gamma^{-s}_+$ and $\Gamma^s_-$ and obtain
\begin{eqnarray}
\left. Z^{ref}_{BPS}\right|_{(R>0, 0<n<B<n+1)}&=&\langle0|\prod_{i=1}^\infty\Gamma^s_+\left(q_{1}^{i-\frac12-\delta_1 }q_2^{-\delta_2}(-Q)^{ \frac12 }\right)\Gamma^{-s}_+\left(q_{1}^{i-\frac12-\delta_1 }q_2^{-\delta_2}(-Q)^{- \frac12 }\right)\nonumber\\
&&\bullet\prod_{j=1}^\infty\Gamma^s_-\left(q_{2}^{j- \frac12+\delta_2 }q_1^{\delta_1}(-Q)^{- \frac12 }\right)\Gamma^{-s}_-\left(q_{2}^{j-\frac12+\delta_2 }q_1^{\delta_1}(-Q)^{ \frac12 }\right)|0\rangle\nonumber\\
&&\bullet\prod _{l=1}^n \prod _{p=1}^l \left(1-Q^{-1} q_1^{n-l+ \frac12 }\text{  }q_2^{p- \frac12 }\right)^{-1}
\end{eqnarray}
Then we insert the matrix model identity operator $\mathbb I$. Thus the corresponding matrix model is
\begin{eqnarray}
  %\label{eq:60}
&&  \left. Z^{ref}_{BPS}\right|_{(R>0, 0<n<B<n+1)}=\int dU\prod _{l=1}^n \prod _{p=1}^l \left(1-Q^{-1} q_1^{n-l+ \frac12 }\text{  }q_2^{p- \frac12 }\right)^{-1}\\
&\times&\left\langle0\left|\prod_{l=1}^\infty\Gamma^s_+\left(q_{1}^{l-\frac12-\delta_1 }q_2^{-\delta_2}(-Q)^{ \frac12 }\right)\Gamma^{-s}_+\left(q_{1}^{l-\frac12-\delta_1 }q_2^{-\delta_2}(-Q)^{- \frac12 }\right)\prod _{i=1}^\infty \Gamma'_-(u_i) \right|0\right\rangle\nonumber\\
&\times&\left\langle 0\left|\prod _{j=1}^\infty \Gamma'_+\left(u_j^{-1}\right)\prod_{r=1}^\infty\Gamma^s_-\left(q_{2}^{r- \frac12+\delta_2 }q_1^{\delta_1}(-Q)^{- \frac12 }\right)\Gamma^{-s}_-\left(q_{2}^{r-\frac12+\delta_2 }q_1^{\delta_1}(-Q)^{ \frac12 }\right)\right|0\right\rangle. \nonumber
\end{eqnarray}
Hence according to the equation (\ref{eq:9}, \ref{eq:20}), the partition function of the matrix model is
\begin{eqnarray}
  %\label{eq:63}
  &&\left. Z^{ref}_{BPS}\right|_{(R>0, 0<n<B<n+1)}=\int dU\prod _{l=1}^n \prod _{p=1}^l \left(1-Q^{-1} q_1^{n-l+ \frac12 }q_2^{p- \frac12 }\right)^{-1}\nonumber\\
  &&\bullet\text{Det}^s\left[\prod _{k=0}^{\infty } \frac{\left(1+s U^{-1}(-Q)^{- \frac12} q_2^{k-\frac12+\delta_2}q_1^{\delta_1}\right)
 \left(1+s U (-Q)^{\frac12 }q_1^{k-\frac12-\delta_1}q_2^{-\delta_2}\right)}{\left(1-sU^{-1}(-Q)^{ \frac12}q_2^{k-\frac12+\delta_2}q_1^{\delta_1}\right)
\left(1-sU (-Q)^{- \frac12} q_1^{k-\frac12-\delta_1}q_2^{-\delta_2}\right)}\right].
\end{eqnarray}

\subsection{The chamber $(R>0, n-1<B<n\leq0)$}

The refined partition function for chamber $(R>0, n-1<B<n\leq0)$  of the resolved conifold is
\begin{eqnarray}
  %\label{eq:59}
\left.Z_{BPS}^{ref}\right|_{(R>0,n-1<B<n\leq0)}&=&  \left\langle 0\left|\prod _{k=1}^{\infty } \Gamma _+^s\left(q_1^{k- \frac12-\delta_1 }q_2^{-\delta_2}(-Q)^{ -\frac12 }\right) \Gamma _+^{-s}
\left(q_1^{k+n- \frac12-\delta_1 }q_2^{-\delta_2}(-Q)^{ \frac12 }\right) \right.\right.\nonumber\\
&&\bullet\prod _{l=1}^n \Gamma _-^s\left(q_2^{l- \frac12+\delta_2 }q_1^{\delta_1}(-Q)^{\frac12 }\right) \Gamma _+^{-s}\left(q_1^{n-l+ \frac12-\delta_1 }q_2^{-\delta_2}(-Q)^{\frac12 }\right)   \nonumber\\
&&\left.\left.\bullet \prod _{k=1}^{\infty}  \Gamma _-^s\left(q_2^{k+n- \frac12+\delta_2 }q_1^{\delta_1}(-Q)^{\frac12 }\right)
\Gamma_-^{-s}\left(q_2^{k- \frac12+\delta_2 }q_1^{\delta_1}(-Q)^{ -\frac12 }\right) \right|0\right\rangle.\nonumber\\
&&
\end{eqnarray}
We use the commutation relation of $\Gamma^{-s}_+$ and $\Gamma^s_-$ and obtain
\begin{eqnarray}
\left. Z^{ref}_{BPS}\right|_{(R>0, n-1<B<n\leq0)}&=&\langle0|\prod_{i=1}^\infty\Gamma^s_+\left(q_{1}^{i-\frac12-\delta_1 }q_2^{-\delta_2}(-Q)^{ -\frac12 }\right)\Gamma^{-s}_+\left(q_{1}^{i-\frac12-\delta_1 }q_2^{-\delta_2}(-Q)^{\frac12 }\right)\nonumber\\
&&\bullet\prod_{j=1}^\infty\Gamma^s_-\left(q_{2}^{j- \frac12+\delta_2 }q_1^{\delta_1}(-Q)^{\frac12 }\right)\Gamma^{-s}_-\left(q_{2}^{j-\frac12+\delta_2 }q_1^{\delta_1}(-Q)^{ -\frac12 }\right)|0\rangle\nonumber\\
&&\bullet\prod _{l=1}^n \prod _{p=1}^l \left(1-Q q_1^{n-l+ \frac12 }\text{  }q_2^{p- \frac12 }\right)^{-1}
\end{eqnarray}
Then we insert the matrix model identity operator $\mathbb I$. Thus the corresponding matrix model is
\begin{eqnarray}
  \label{eq:60}
&&  \left. Z^{ref}_{BPS}\right|_{(R>0, n-1<B<n\leq0)}=\int dU\prod _{l=1}^n \prod _{p=1}^l \left(1-Q q_1^{n-l+ \frac12 }\text{  }q_2^{p- \frac12 }\right)^{-1}\\
&\times&\left\langle0\left|\prod_{l=1}^\infty\Gamma^s_+\left(q_{1}^{l-\frac12-\delta_1 }q_2^{-\delta_2}(-Q)^{ -\frac12 }\right)\Gamma^{-s}_+\left(q_{1}^{l-\frac12-\delta_1 }q_2^{-\delta_2}(-Q)^{ \frac12 }\right)\prod _{i=1}^\infty \Gamma'_-(u_i) \right|0\right\rangle\nonumber\\
&\times&\left\langle 0\left|\prod _{j=1}^\infty \Gamma'_+\left(u_j^{-1}\right)\prod_{r=1}^\infty\Gamma^s_-\left(q_{2}^{r- \frac12+\delta_2 }q_1^{\delta_1}(-Q)^{\frac12 }\right)\Gamma^{-s}_-\left(q_{2}^{r-\frac12+\delta_2 }q_1^{\delta_1}(-Q)^{-\frac12 }\right)\right|0\right\rangle. \nonumber
\end{eqnarray}
Hence according to the equation (\ref{eq:9}, \ref{eq:20}), the partition function of the matrix model is
\begin{eqnarray}
  \label{eq:63}
  &&\left. Z^{ref}_{BPS}\right|_{(R>0, n-1<B<n\leq0)}=\int dU\prod _{l=1}^n \prod _{p=1}^l \left(1-Q q_1^{n-l+ \frac12 }q_2^{p- \frac12 }\right)^{-1}\nonumber\\
  &&\bullet\text{Det}^s\left[\prod _{k=0}^{\infty } \frac{\left(1+s U^{-1}(-Q)^{\frac12} q_2^{r-\frac12+\delta_2}q_1^{\delta_1}\right)
 \left(1+s U (-Q)^{-\frac12 }q_1^{l-\frac12-\delta_1}q_2^{-\delta_2}\right)}{\left(1-sU^{-1}(-Q)^{ -\frac12}q_2^{r-\frac12+\delta_2}q_1^{\delta_1}\right)
\left(1-sU (-Q)^{ \frac12} q_1^{l-\frac12-\delta_1}q_2^{-\delta_2}\right)}\right].
\end{eqnarray}

%\subsection{The chamber $(R<0, 0<n<B<n+1)$}

%According to the results in \cite{Liu2010}, the refined BPS states partition function in the chamber $(R<0, 0<n<B<n+1)$ is
%\begin{eqnarray}
%Z^{ref}_{BPS}|_{(R<0, 0<n<B<n+1)}&=&\langle0|\Gamma_+^{-s}\left[ q_1^{\frac{1}{2}-\delta_1}q_2^{-\delta_2}(-Q)^{-\frac{1}{2}}\right]\Gamma_-^s\left[q_2^{n-\frac{1}{2}+\delta_2}q_1^{\delta_1}(-Q)^{-\frac{1}{2}}\right]\cdots\nonumber\\
%&&\bullet\Gamma_+^{-s}\left[ q_1^{n-\frac{1}{2}-\delta_1}q_2^{-\delta_2}(-Q)^{-\frac{1}{2}}\right] \Gamma_-\left[q_2^{\frac{1}{2}+\delta_2}q_1^{\delta_1}(-Q)^{-\frac{1}{2}}\right]|0\rangle\nonumber\\
%&=& (1-q_1^{\frac{1}{2}}q_2^{n-\frac{1}{2}}Q^{-1})\cdots (1-q_1^{n-\frac{1}{2}}q_2^{\frac{1}{2}}Q^{-1}).
%\end{eqnarray}

\bibliographystyle{JHEP}
\bibliography{RWC}
\end{document}